\documentclass[prd,aps,nofootinbib,superscriptaddress,floatfix,preprintnumbers,twocolumn]{revtex4}
\usepackage[utf8]{inputenc}
\usepackage{amsfonts}
\usepackage{hyperref,amssymb,amsmath,graphicx,xcolor,slashed}
\usepackage[acronym]{glossaries}
\newacronym[longplural=parton distribution functions]{pdf}{PDF}{parton distribution function}
\newacronym{lamet}{LaMET}{large momentum effective theory}
\newacronym{qcd}{QCD}{quantum chromodynamics}

\makeglossaries
\DeclareUnicodeCharacter{00A0}{ }

\begin{document}
\title{Leading Power Accuracy in Lattice Calculations of Parton Distributions}
\author{Rui Zhang}
\email{rayzhang@umd.edu}
\affiliation{Department of Physics, University of Maryland, College Park, MD 20742, USA}
\author{Jack Holligan}
\email{holligan@msu.edu}
\affiliation{Biomedical and Physical Sciences Building, Michigan State University, East Lansing, MI, 48824, USA}
\author{Xiangdong Ji}
\email{xji@umd.edu}
\affiliation{Department of Physics, University of Maryland, College Park, MD 20742, USA}
\author{Yushan Su}
\email{ysu12345@umd.edu}
\affiliation{Department of Physics, University of Maryland, College Park, MD 20742, USA}
\affiliation{Physics Division, Argonne National Laboratory, Lemont, Illinois 60439, USA}
\begin{abstract}

In lattice-QCD calculations of parton distribution functions (PDFs) via large-momentum effective theory, the leading power (twist-three) correction appears as ${\cal O}(\Lambda_{\rm QCD}/P^z)$ due to the linear-divergent self-energy of Wilson line in quasi-PDF operators. For lattice data with hadron momentum $P^z$ of a few GeV, this correction is dominant in matching, as large as 30\% or more. We show how to eliminate this uncertainty through choosing the mass renormalization parameter consistently with the resummation scheme of the infrared-renormalon series in perturbative matching coefficients. An example on the lattice pion PDF data at $P^z = 1.9$ GeV shows an improvement of matching accuracy by a factor of more than $3\sim 5$ in the expansion region $x= 0.2\sim 0.5$.

\end{abstract}

\maketitle

{\it Introduction~~}  The parton distribution functions (PDFs) of the hadrons, describing the momentum distribution of quarks and gluons inside, serve as key inputs in theoretical predictions of the hadronic cross sections in high-energy scattering
experiments~\cite{Ellis:1996mzs}. They have been determined to high precision by global fittings to experimental data~\cite{Aicher:2010cb,Ball:2017nwa,Hou:2019efy,Bailey:2020ooq,Novikov:2020snp,Barry:2021osv}. At present, a direct way to calculate the Bjorken-$x$-dependent PDFs non-perturbatively is through lattice QCD simulations of Euclidean correlation functions, interpreted, e.g., in the framework of the large momentum expansion or large-momentum effective theory (LaMET)~\cite{Ji:2013dva,Ji:2014gla}. A recent summary of PDF lattice
calculations with various other approaches can be found in Ref.~\cite{Constantinou:2022yye}.

Since the proposal of LaMET, much progress has been made within the framework in first-principle predictions of PDFs~\cite{Constantinou:2020hdm,Ji:2020ect,Constantinou:2022yye}. The calculations in large-momentum expansion have been improved to a level that the precision control at various steps has become important~\cite{Gao:2021dbh}. Apart from the usual requirements of lattice QCD, such as physical pion mass, large volume and small lattice spacing, one needs to properly renormalize the lattice data to achieve a reliable continuum limit and matching to physical PDFs with high precision. Since the LaMET expansion parameter is both in coupling $\alpha_s(\sim P^z)$ and in powers of $\Lambda_{\rm QCD}/P^z$, and since most of the calculations so far are limited to $P^z\sim $ few GeV, it is critically important to understand the effects of power corrections, which have not been properly addressed in the literature so far. In a recent paper~\cite{Su:2022fiu}, it has been shown that the proper perturbative scale in LaMET matching is around $2xP^z$, which implies that the power correction of type $\Lambda_{\rm QCD}/2xP^z$ can reach $\sim$ 30\% level at lower $x$ limit of 0.2, and as large as 15-20\% at $x=0.5$. Therefore, achieving twist-three power accuracy is necessary for a precision prediction of PDFs in the lower $x$ region. Once done, the naive counting puts the twist-four power correction to order $(\Lambda_{\rm QCD}/2xP^z)^2$ which is below 10\% in the LaMET expansion region. 

In the matching formula for PDFs, there are two sources of the twist-three power corrections. First, the linearly-divergent self-energy of Wilson line has an intrinsic ambiguity $\mathcal{O}(\Lambda_{\rm QCD})$ due to long-range non-perturbative effects~\cite{Beneke:1992ea,Beneke:1992ch,Beneke:1998ui,Bauer:2011ws,Bali:2013pla,Bali:2014fea,Bali:2014sja,Ayala:2019uaw,Ayala:2020pxq} and, corresponding to the freedom of a finite term for subtraction in renormalization. Different choices introduce a scheme dependence through a twist-three mass parameter $m_0$ in the form of a $e^{m_0z}$ factor in quasi-PDFs, corresponding to a $\mathcal{O}(\Lambda_{\rm QCD}/xP^z)$ freedom in the momentum space quasi-PDF~\cite{Ji:2020brr}. Meanwhile, the perturbative matching kernel to the light-cone PDFs is an asymptotic QCD perturbation series in $\alpha_s$ with the coefficients grow factorially at high orders~\cite{Braun:2018brg}, a phenomenon called infrared renormalons~\cite{tHooft:1977}. The leading renormalon series can be summed/regularized in multiple ways that differ by a term linear in the spatial Wilson line length $z$, also introducing a twist-three scheme dependence. Leading power accuracy is achieved by choosing both renormalization and matching schemes consistently, thereby removing $\mathcal{O}(\Lambda_{\rm QCD}/xP^z)$ twist-three contribution entirely in the light-cone PDF extraction.

In this paper, we show how to achieve twist-three power accuracy in calculations of PDFs for the first time. We use perturbative matching coefficients accurate to next-to-next-to leading order (NNLO) plus the leading renormalon contribution estimated from the heavy quark pole mass calculations~\cite{Pineda:2001zq,Ayala:2014yxa}. The results of the latter at $n_f=0$ agree well with a precise lattice heavy-mass perturbative series up to $~20$th order in the QCD coupling~\cite{Bali:2013pla}. We adopt the renormalization condition that the continuum-limit lattice matrix elements have a short-distance ($z$) expansion consistent with Wilson coefficients resummed through the principal value (PV) prescription. Using this, we extract the finite renormalization or twist-three mass parameter $m_0$ from the $P^z=0$ lattice matrix element in the perturbative $z$ region~\cite{LatticePartonCollaborationLPC:2021xdx}. The scale-independence of $m_0$ and fast convergence between NLO and NNLO perturbative calculations are important indicators that leading power contribution is under control. As a demonstrative example, we show that the uncertainty in $x$-dependent pion PDF is significantly reduced after the leading power correction. 

\vskip 0.2cm
{\it Linear divergence and uncertainty in twist-three contribution~~~} Most LaMET calculations start from the coordinate space bare correlator~\cite{Ji:2013dva}
\begin{align}
    h^B(z,P_z,a) = \langle P^z |\bar{\psi}(z)\Gamma W(z,0)\psi(0)|P^z\rangle
\end{align}
where $W(z,0)=\mathcal{P}\exp[\int_0^zA_z(z')dz']$ with path ordering is a gauge link along the $z$-direction, making the operator gauge invariant. Due to the self-energy of the Wilson line, the above operator has a linear divergence which can be renormalized in the multiplicative factor of $e^{-\delta m(a)z}$, equivalent to the quantum correction to the mass $\delta m(a)$ of an auxiliary ``infinitely-heavy quark'' (whose propagator $Q(z)\bar{Q}(0)$ generates the Wilson line $W(z,0)$~\cite{Chen:2016fxx,Green:2017xeu,Ji:2017oey,Ishikawa:2017faj,Ji:2020brr}). However, the quark carries a color charge, which leads to a ill-defined infrared contribution to the mass. Thus, one can choose the renormalization factor either with a ``short-distance mass" defined by an infrared-regulated ``pole mass''~\cite{Bali:2013pla}, or with a non-perturbatively defined mass parameter in terms of a physical matrix element~\cite{Zhang:2017bzy,LatticePartonCollaborationLPC:2021xdx,Gao:2021dbh}.  Different mass subtractions (labelled by a parameter $\tau$) differ by the intrinsic non-perturbative scale, $\Delta \left(\delta m(a,\tau)\right)\sim \Lambda_{\rm QCD}$. Thus, besides the usual renormalization scale $\mu$ for logarithmic divergences, 
the quasi-PDF matrix element has the $\tau$ dependence from the renormalization of the linear divergence.

To achieve high precision, it has been suggested to eliminate the power-divergence from lattice data through a self-renormalization approach~\cite{LatticePartonCollaborationLPC:2021xdx}.  The lattice ultra-violet (UV) regulator $a$ dependence of the $P_z=0$ quasi-PDF matrix elements $h^B(z,a,P_z=0)$ has been parameterized by fitting the lattice data on multiple lattice spacings. Multiple sets of parameters were found corresponding to the same minimal $\chi^2$ fit, can be regarded as different choices of $\tau$. These different sets show an $\mathcal{O}(\Lambda_{\rm QCD})$ ambiguity in the mass renormalization $\delta m(a,\tau)$, an example showing the $\tau$-scheme dependence. 

The $\tau$-dependence entails that the short-distance ($z$) expansion~\cite{Izubuchi:2018srq} of the renormalized correlator must contain a twist-three non-perturbative parameter $m_0(\tau)$ of order $\Lambda_{\rm QCD}$ contributing as a linear-$z$ term (the sign is chosen so $m_0$ contributes positively as ``residual mass''), 
\begin{align}
\label{eq:ope}
& h^R(z,P_z,\mu,\tau) \\ &= \Big(1-m_0(\tau) z\Big)\sum_{k=0}^\infty C_k\left(\alpha_s(\mu),\mu^2 z^2\right)
\lambda^k a_{k+1}(\mu)+ {\cal O}(z^2)\nonumber \\ 
&=\sum_{k=0}^\infty \Big[C_k\left(\alpha_s(\mu),\mu^2 z^2\right) - zm_0 (\tau) \Big] \lambda^k a_{k+1} (\mu) +  {\cal O}(z\alpha_s, z^2), \nonumber
\end{align}
where $\lambda = zP^z$,  $a_k$ are spin-$k$ twist-2 matrix elements, and perturbative series $C_k$ are the associated Wilson coefficients. The twist-three contribution is universal in the sense that it multiplies the leading-twist term in the same manner independent of the spin of the local operators. On the other hand, $m_0(\tau)$ does depend on the external states in which the correlator matrix elements are taken as well as kinematic Lorentz invariants formed by external momenta. In the LaMET expansion, the above twist-three term translates to a linear one in inverse large momentum $1/P_z$. 

To extract the twist-two PDFs up to  
leading-power accuracy, one needs to fix $m_0(\tau)$ from a non-perturbative matrix element. 
A natural choice is the $P_z=0$ case where only $k=0$ term in Eq.~\eqref{eq:ope} survives when $\lambda=0$. With renormalized lattice data  $h^R(z,P_z=0, \mu,\tau)$, we can fit $m_0(\tau)$ with the expansion at linear-$z$ accuracy using known $a_1=1$ and $\overline{\rm MS}$ series $C_0(\alpha_s(\mu),\mu^2z^2)$. 

The fitted $m_0(\tau)$ can alternatively be absorbed into an additional finite renormalization to define a new $\tau$-independent renormalizaiton factor $Z^R(z,a,\mu)\sim \exp[\delta m(a,\tau)z+ m_0(\tau)z]$ and the renormalized quasi-PDF $\tilde{q}(x,P_z,\mu)$ for $P_z>0$, the latter can be used in the standard LaMET matching up to leading power accuracy but without the explicit twist-three contribution. 

The determination of $m_0(\tau)$ requires a rigorous relation between the bare lattice data $h^B(z,P_z,a^{-1})$ and the renormalized matrix elements $h^R(z,P_z,\mu,\tau)$ in $\overline{\rm MS}$ scheme, or equivalently, the rigorous definition of the $P_z$-independent renormalization constant $Z^R(z,a,\mu,\tau)$,
\begin{align}
    h^R(z,P_z,\mu,\tau)=h^B(z,P_z,a^{-1})/Z^R(z,a,\mu,\tau).
\end{align}

We start from $h^R(z,P_z=0,\mu=z^{-1})$ at small $z$ in $\overline{\rm MS}$ scheme without regularizing the twist-3 ambiguity, and solve the renormalization group (RG) equation to separate the dependence on $z$ and the UV regulator $a^{-1}$ (or UV renormalization scale $\mu$ in dimensional regularization), 
\begin{equation}
    \frac{\partial h^R(z,0,\mu)}{\partial\ln\mu^2} = \gamma(\mu) h^R(z,0,\mu) \ , 
\end{equation}
where the anomalous dimension $\gamma(\alpha(\mu))$ of $h^R(z,0, \mu)$ has been calculated up to 3-loop order for the quasi-PDF operator~\cite{Braun:2020ymy}. 
The above equation can be solved when $z$ is 
perturbative by evolving from the initial point $\mu=z^{-1}$. The solution at scale $\mu$ is then
\begin{align}
    h^R(z,0,\mu)
    =&h^R (z,0,z^{-1})\exp{\left(\int_{\alpha(z^{-1})}^{\alpha(\mu)}\frac{\gamma(\alpha')}{\beta(\alpha')}d\alpha'\right)}\nonumber\\
    =&h^R(z,0,z^{-1})\exp{\left[-\mathcal{I}(z^{-1})\right]}\exp{\left[\mathcal{I}(\mu)\right]}.
    \label{eq:me_resum}
\end{align}
where $\mathcal{I}(\mu)$ is a analytic function of $\alpha(\mu)$ because both $\gamma(\alpha)$ and $\beta(\alpha)$ are polynomials of $\alpha$ when truncated at a certain order in perturbation theory.
Similarly, for the lattice scheme with explicit linear divergence, the matrix elements now have the following form:
\begin{align}
    \!\!h^B(z,0,a^{-1})=e^{-\delta m(a) z}h^{\rm lat}(z,0,z^{-1})e^{-\mathcal{I^{{\rm lat}}}(z^{-1})}e^{\mathcal{I^{{\rm lat}}}(a^{-1})}
\end{align}
where $h^{{\rm lat}}$ is the perturbation series evaluated at scale $z^{-1}$ in the lattice scheme. In both schemes, the $z$-dependence is now completely factorized. In the lattice matrix element, we have manifestly separated out the mass dependence with the linearly-divergent mass correction $\delta m(a)$. The $z$-dependence is physical, thus should not depend on scheme choices. So we can identify $h^{\rm lat}(z,0,z^{-1})e^{-\mathcal{I^{{\rm lat}}}(z^{-1})}$ in the lattice scheme and $h(z,0,z^{-1})\exp{\left[-\mathcal{I}(z^{-1})\right]}$ in the $\overline{\rm MS}$ scheme.
The above expressions allow us to express the lattice matrix element in terms of the $\overline{\rm MS}$ perturbation series 
\begin{align}
\label{eq:master_2}
    &h^B(z,0,a^{-1})=h^R(z,0,z^{-1})e^{\mathcal{-I}(z^{-1})}e^{\mathcal{I^{{\rm lat}}}(a^{-1})}e^{-\delta m(a) z},
\end{align}
when combined with Eq.~\eqref{eq:me_resum}, indicating that a twist-3 accurate renormalization constant can be defined as
\begin{align}
\label{eq:renorm}
    Z^R(z,a,\mu,\tau)=\exp[{\mathcal{I}^{{\rm lat}}(a^{-1})-\mathcal{I}(\mu)-\delta m (a,\tau) z-m_0(\tau)z}],
\end{align}
which also works for $P_z>0$ lattice data. Now the mass correction $\delta m (a,\tau)$ is fixed in the $\tau$-scheme, and a corresponding non-perturbative mass parameter $m_0(\tau)$ in Eq.~\eqref{eq:ope} is absorbed into the renormalization to guarantee the twist-3 accuracy.
After a rearrangement, Eq.~\eqref{eq:master_2} becomes
\begin{align}
\label{eq:master_3}
   m_0(\tau)z&(1+\mathcal{O}(z\Lambda_{\rm QCD}))=\ln\left[h^R(z,0,z^{-1},\tau)e^{\mathcal{-I}(z^{-1})}\right]\nonumber\\
   -&\ln \left[h^{B}(z,0,a^{-1})e^{-\mathcal{I}^{{\rm lat}}(a^{-1})}e^{\delta m(a,\tau) z}\right],
\end{align}
that allows us to extract $m_0(\tau)$ as a slope of such a quantity.

\begin{figure}[tbhp!]
    \centering
    \includegraphics[width=0.9\linewidth]{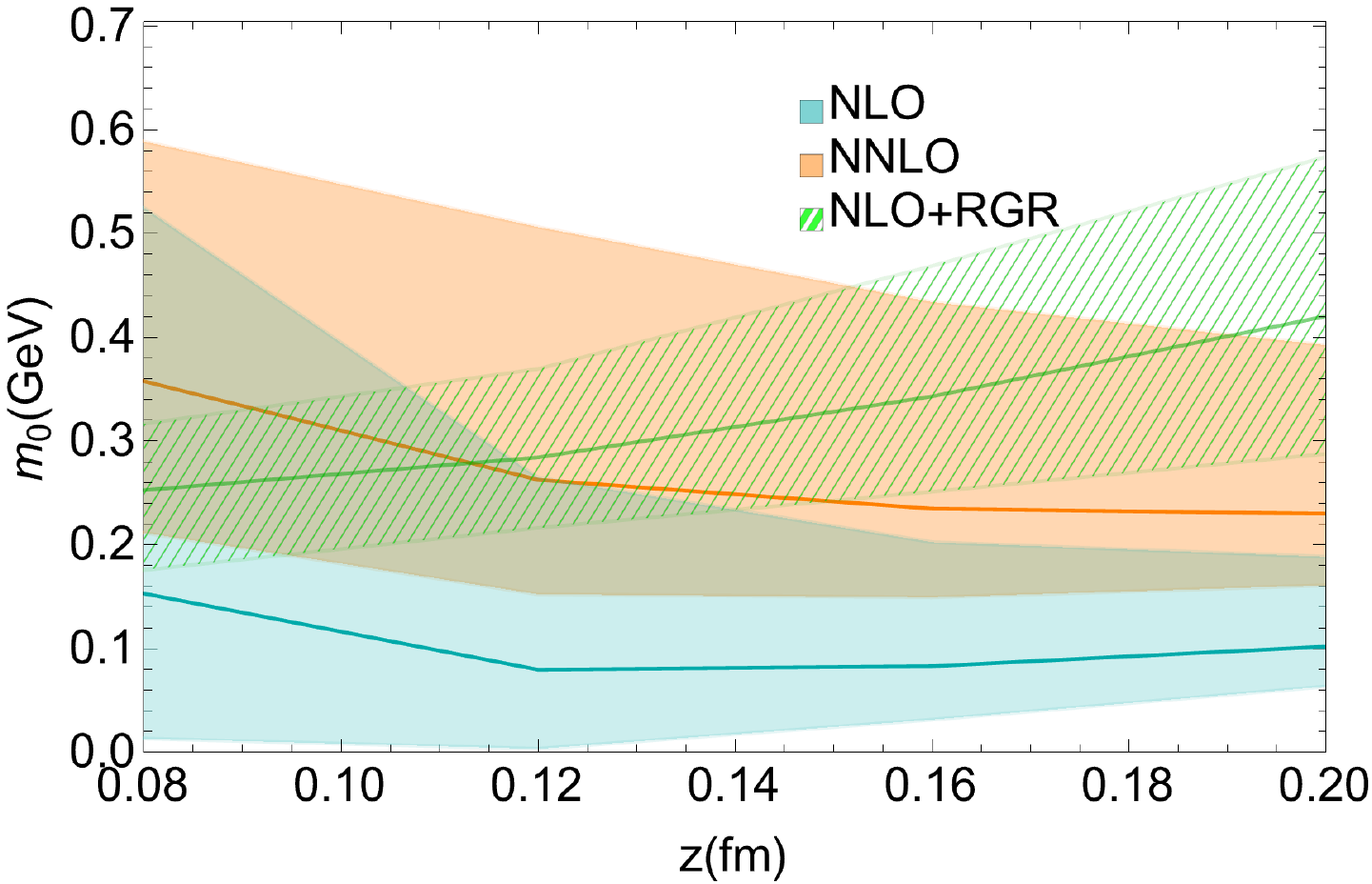}
    \includegraphics[width=0.9\linewidth]{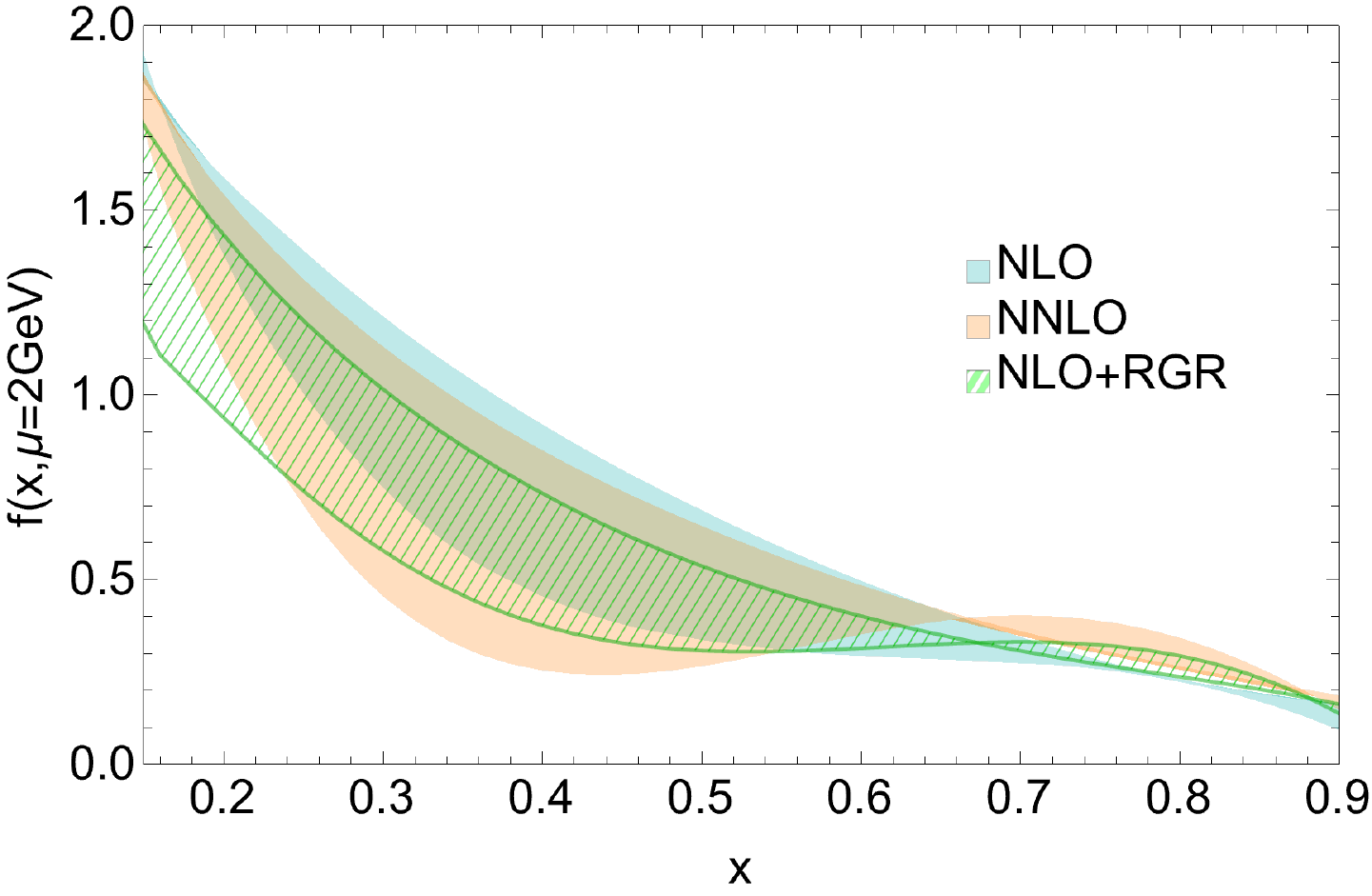}
    \caption{Top: Uncertainty in $m_0(\tau)$ extracted from the pion $P_z=0$ matrix element and fixed-order Wilson coefficients with and without RG resummation. The band width shows the renormalizaiton scale $\mu$ dependence.  Bottom: The uncertainties in the pion light-cone PDF extracted from LaMET expansion with the above $m_0$. The overlap region exhibits a darker color.}
    \label{fig:nlo_compare}
\end{figure}

In Ref.~\cite{LatticePartonCollaborationLPC:2021xdx}, the fixed (next-to-leading, NLO) order Wilson coefficient~\cite{Izubuchi:2018srq,Li:2020xml}
\begin{align}
    C_0(\alpha_s(\mu),\mu^2z^2)=1+\frac{\alpha_s(\mu)}{3\pi}\left(3\ln\left(\frac{z^2\mu^2e^{2\gamma_E}}{4}\right)+5\right)
\end{align}
has been used to fit $m_0(\tau)$. The use of this fixed-order expression in Eq.~\eqref{eq:master_3} will in principle introduce small deviations from unresummed $\alpha^n_s(\mu)\ln^n(z^2\mu^2)$ terms thus slightly alters the $z$-dependence. The result will then depend on the choice of both $\mu$ and $z$. 
Here we first follow their strategy and examine the uncertainty in this fitting method. For a single lattice spacing, $I_0=e^{-\mathcal{I}^{{\rm lat}}(a^{-1})}$ is a constant thus is absorbed into the interception and not affecting the slope $m_0(\tau)$, when fitted to an approximate form of Eq.~\eqref{eq:master_3} at fixed order perturbation theory
\begin{align}
\label{eq:master_4}
   m_0(\tau)z-I_0&\approx\ln\left[e^{-\delta m(a,\tau) z}h^R(z,0,\mu,\tau)/h^{B}(z,0,a^{-1})\right].
\end{align} 
We fit $m_0(\tau)$ to Eq.~\eqref{eq:master_4} in a range $[z-a,z+a]$ from the pion quasi-PDF lattice data from the ANL and BNL collaboration~\cite{Gao:2021dbh} with lattice spacing $a=0.04$~fm. 
As shown by cyan and orange bands in the top panel in Fig.~\ref{fig:nlo_compare}, the fixed-order (NLO or NNLO) $C_0$ introduces a large uncertainty from scale $\mu$ variation from $1$ GeV to $4$ GeV, corresponding to all possible relevant physics scales ($2xP^z$) in the problem. We can resum the large logarithmic terms $\alpha^n_s(\mu)\ln^n(z^2\mu^2)$ in $C_0$ to reduce the $\mu$ dependence using the renormalization group resummed (RGR) renormalization constant defined in Eq.~\eqref{eq:renorm}. We then fit $m_0$ to Eq.~\eqref{eq:master_3}, shown as the hatched green band on the top panel, has a strong dependence on $z$, and becomes unreasonably large at larger $z>$ 0.2~fm. This is a indication that there is a significant contamination effect from unaccounted higher-order terms in twist-two $C_0$, which has logarithmic dependence in $z$. The $z$-dependence is altered by the truncation in $\alpha_s(z^{-1})$. Note that the uncertainty shown is purely systematic error, because the data are so clean at $P_z=0$ and $z<0.2$~fm that can be completely neglected in the extraction of $m_0$.

This large uncertainty in the twist-three mass parameter $m_0$ can be translated to that for light-cone PDFs in LaMET calculations. The extracted isovector light-cone PDF is shown in the bottom plot of Fig.~\ref{fig:nlo_compare}. With the fixed order $C_k$, the uncertainty in $m_0$ yields $> 30\%$ error near $x\sim 0.3$. With renormalization group improvement, a significant uncertainty in the extracted PDF still exists, shown as the hatched green band with NLO+RGR label in the same plot. These large uncertainties indicate that improving calculations up to twist-three accuracy is crucially important for lattice data at $P^z\sim 2$ GeV. 

\vskip 0.2cm

{\it Power accuracy in matching coefficients through infrared renormalons~~} It is well-known that perturbative Wilson
coefficients $C_k(\alpha_s)=\sum_n c_{kn}\alpha_s^n$ are asymptotic series because of infra-red contributions at large orders, a phenomenon called infrared renormalons (IRR) ~\cite{tHooft:1977}. For the quasi-PDF operator, the leading IRR comes from the long-distance contributions to the 
self-energy of the charged Wilson line, and resulting $c_{kn}$ growing factorially $\propto n!(\beta_0/2\pi)^n$ at large-$n$ orders~\cite{Braun:2018brg}, where $\beta_0 =11-2n_f/3$ ($n_f$ is the number of active quark flavors) is the first term of the QCD beta function. This behavior is exactly the same as the perturbation series for the ``pole" mass of a quark~\cite{Pineda:2001zq,Bali:2013pla}.  

The perturbative runaway infrared contributions at large orders must be regularized, or equivalently, one has to specify a way to resum the perturbative series. Different methods of resummation/regularization 
yield results differing by the order of the minimal term in the series, at $n\sim 2\pi/\beta_0\alpha_s$~\cite{Beneke:1992ch}.
A simple estimation shows that this is twist-three contribution ${\cal O}(z\Lambda_{\rm QCD})$, a linear power in $z$ (other renormalon poles corresponding to higher-power-$z$/twist terms). Therefore, the twist-two contributions themselves are ambiguous up to higher-twist contributions, and the twist-three parameter $m_0(\tau)$ depends on the resummation/regularization method for the leading renormalon series in $C_k(\alpha_s)$. 
There have been a number of proposals for renormalon regularization in the literature~\cite{Beneke:1998ui}. For example, one can define subtraction for infrared contributions at every order in perturbation series. One can also 
calculate the series in usual $\overline{\rm MS}$ or lattice method and define an all-order sum through a Borel transformation. In this regard, it has been advocated to use the principal value prescription to regulate the renormalon poles in the Borel integral~\cite{Ayala:2019uaw,Ayala:2020pxq}. We consider a prescription for $C_k(\alpha_s)$ together with the UV regularization of the correlator as the complete $\tau$-scheme.

Calculating $C_{k}$ to large orders is extremely challenging if not impossible. The only type of diagrams we can calculate to all orders analytically is the bubble-chain ones, which are leading in the large-$n_f$ limit in Abelian gauge theory. When restored to the full $\beta$ function in non-Abelian theories, the leading asymptotic coefficient is $ n!(\beta_0/2\pi)^n$, which is indeed a renormalon corresponding to the quark pole mass.  On the other hand, the actual overall strength (represented by an overall factor $N_m$) of the renormalon 
can be quite different from the bubble-chain diagram result,
$N_m^{\rm bubble~chain}=0.98$~\cite{Beneke:1994rs}. 
Fortunately, lattice numerical calculations to perturbative large orders have become possible for very limited quantities~\cite{Bali:2013pla,Bali:2014fea}. In particular, the pole mass on pure gauge ensembles ($n_f=0$) has been calculated to $n\sim 20$, $m = \mu \sum_n r_n\alpha_s^{n+1}$, and confirmed the conjecture on the leading IR renormalon form at large $n$~\cite{Bali:2013pla}, 
\begin{align}
    r_n = N_m \left(\frac{\beta_0}{2\pi}\right)^n
    \frac{\Gamma(n+1+b)}{\Gamma(1+b)}
    \left[1+\frac{c_1b}{b+n} +...\right],
\end{align} 
where $b=\beta_1/2\beta_0^2$ and $c_1=(\beta_1^2-\beta_0\beta_2)/(4b\beta^4_0)$ are from higher orders in the QCD beta function. Moreover, 
the mass renormalon strength has been determined to be $N_m(n_f=0)=0.660(56)$ in $\overline{\rm MS}$ scheme. Furthermore, using an analytical method 
in Ref. \cite{Pineda:2001zq}, one can estimate $N_m(n_f=0)=0.622$ consistent with the previous determination, and with dynamic fermions, we used $N_m(n_f=3) =0.575$ in this work. 

Using the above state-of-art knowledge on the mass renormalon, we have the leading renormalon contribution for $C_k$ after a Borel transformation,  
\begin{align}
\label{eq:borel_integral}
    &C_k(\alpha_s(z^{-1}),1)_{\rm PV}= N_m\frac{4\pi}{\beta_0}   \int_{\rm 0, PV}^{\infty}du \nonumber \\
  & \times  e^{-\frac{4\pi u}{\alpha_s(z^{-1})\beta_0}}  \frac{1}{(1-2u)^{1+b}}\big(1+c_1(1-2u)+...\big),  
\end{align}
where we have chosen a PV prescription for regulating poles on the Borel $u$-plane. 
We define a leading renormalon resummation (LRR) result by resumming the leading divergent contributions to all orders at $\mu=z^{-1}$, while keeping the lower-order expansion of $C^{\rm LRR}(
\alpha_s)$ the same as $C_0(\alpha_s(\mu),z^2\mu^2)$,
\begin{align}
	C^{\rm LRR}&(\alpha_s(z^{-1}),1)=C_k(\alpha_s(z^{-1}),1)+\nonumber\\
	&\left[ C_k(\alpha_s(z^{-1}),1)_{\rm PV}-\sum_i\alpha^{i+1}_s(z^{-1}) r_i\right].
\end{align}
Now $C^{\rm LRR}(\alpha_s)$ contains not only the fixed-order results calculated explicitly, but also higher-order (twist-two) perturbative terms contributing to the leading factorial growth. 

\begin{figure}[tbhp!]
    \centering
    \includegraphics[width=0.9\linewidth]{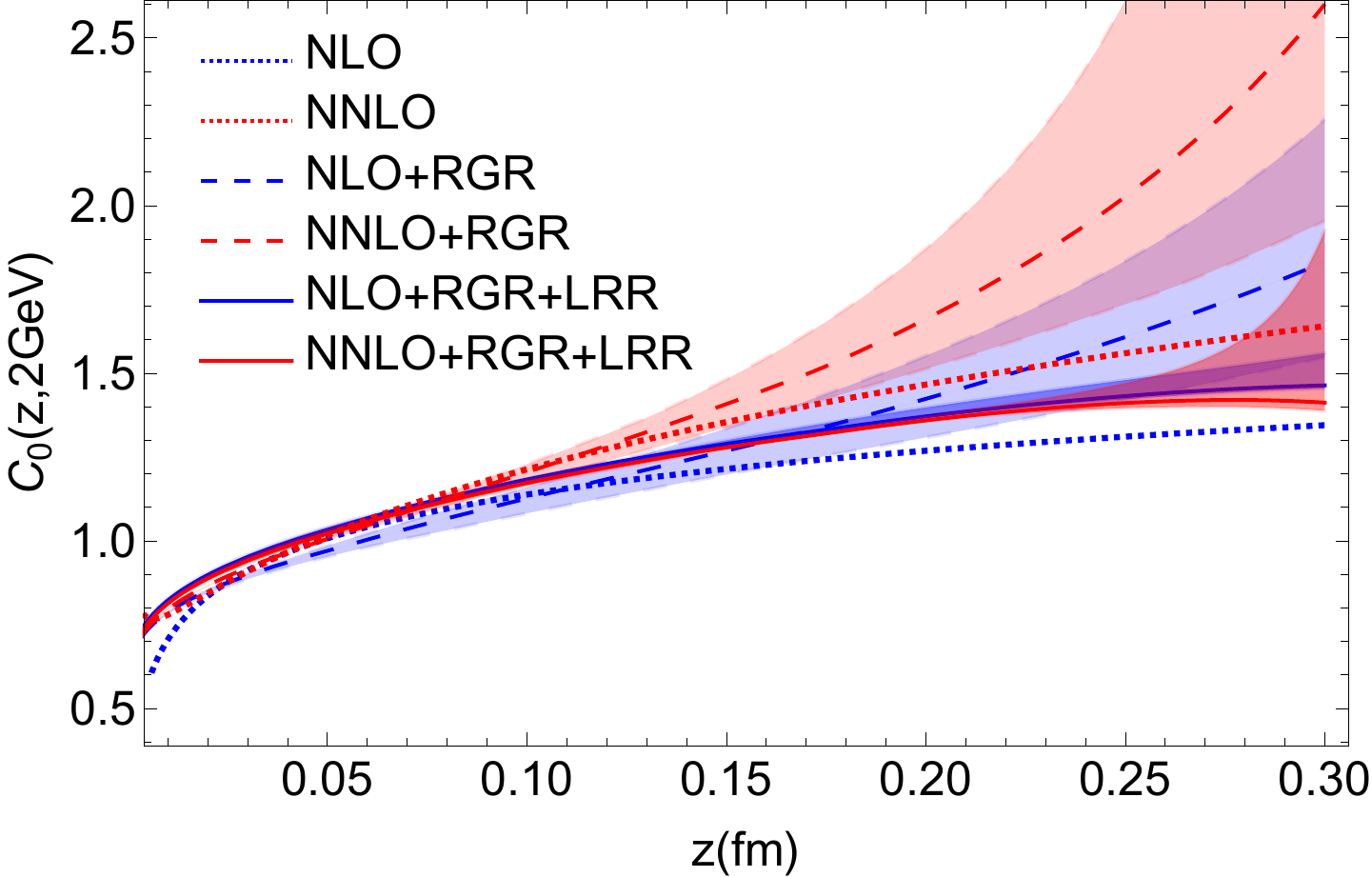}
        \includegraphics[width=0.9\linewidth]{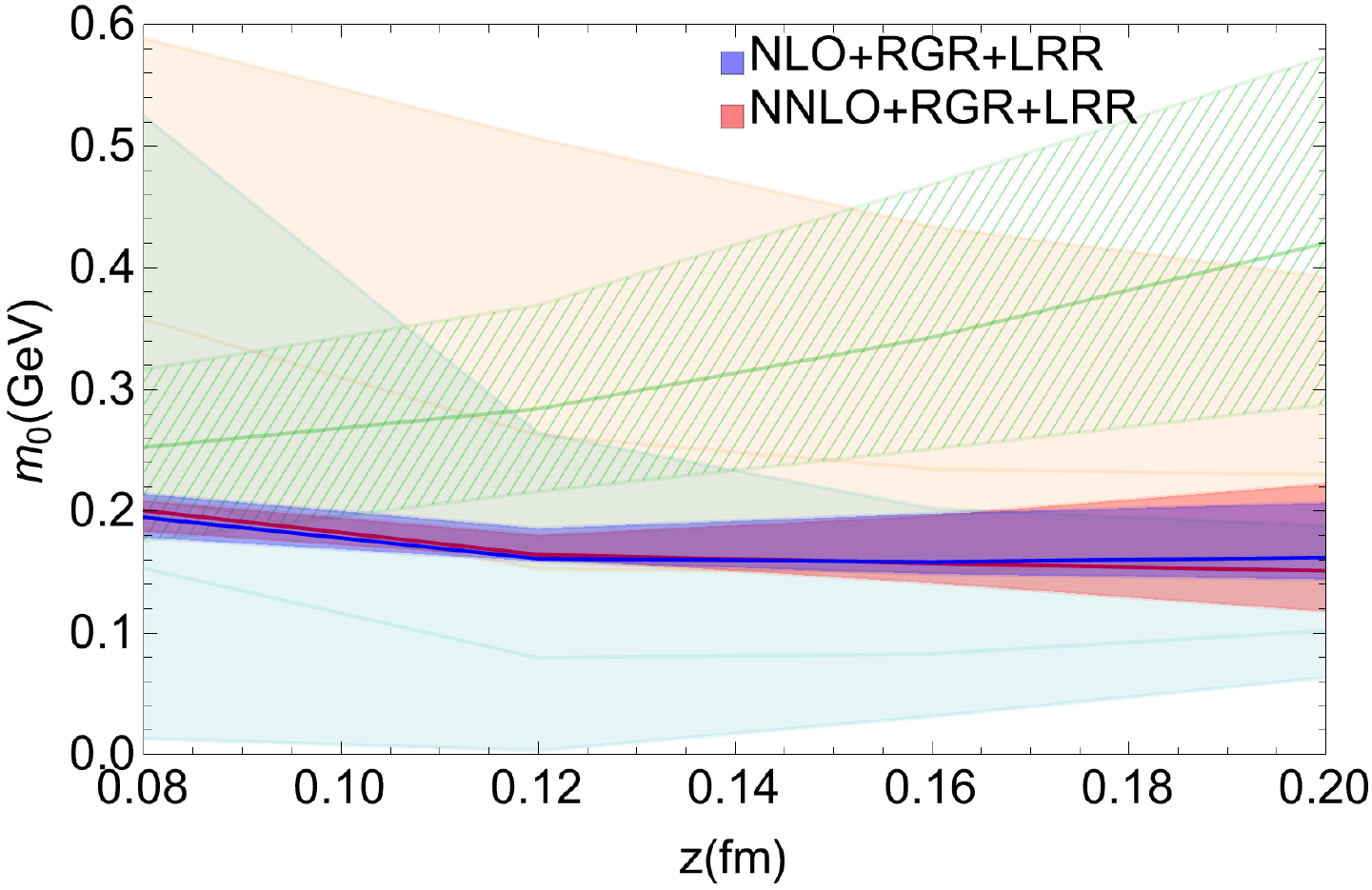}
    \caption{Top: The comparison of $C_0(\alpha_s(\mu),z^2\mu^2)$ from the fixed-order (dotted), renormalization group resummation (dashed), and the leading renormalon resummation (solid). Bottom: $m_0(\tau)$ extracted from leading renormalon resummation with PV as a IR regulator.}
    \label{fig:c0_compare}
\end{figure}

We compare the $k=0$ Wilson coefficient $C_0(\alpha_s(\mu),z^2\mu^2)$ for the fixed-order (NLO+NNLO), fixed order (NLO+NNLO) with RGR, and the LRR-improved formalism on the top panel in Fig.~\ref{fig:c0_compare}. 
The error bands in RGR are obtained by varying the resummation scale from $0.75z^{-1}$ to $1.5z^{-1}$, corresponding to about 30\% change in the strong coupling. Going beyond the lower bound in our data range, the perturbation theory breaks down. While there is a large difference from NLO to NNLO in fixed-order calculations with or without renormalization group improvement, the LRR results show much better convergence in the perturbative region $z<0.3$~fm, and much smaller dependence on the scale variation, indicating that NNLO term is already dominated by the leading renormalon. 

We show the NLO LRR-improved $m_0$ result as blue band together with fixed-order results on the lower panel in Fig.~\ref{fig:c0_compare}. 
By including the leading renormalon, there is now a clear window near $z=0.12$~fm for a constant $m_0(\tau)= 0.161^{+0.025}_{-0.002}$ GeV for NLO with much smaller uncertainty. Thus Eq.~\eqref{eq:ope} achieves the linear-$z$ accuracy when the leading renormalon series is resummed. We also show the NNLO renormalon-resummed results $m_0(\tau)= 0.164^{+0.016}_{-0.003}$ GeV as the red band to demonstrate the good convergence with this method, consistent with the blue band and has smaller scale dependence at small $z$. The difference between the non-perturbative lattice result and the perturbation series
is well described by the linear dependence in $z$ in the perturbatively-reliable region. This gives us 
confidence that we have reached twist-three power accuracy for
describing the $P_z=0$ matrix element.  

\vskip 2mm

{\it PDF matching to leading power accuracy~~}
We commented after Eq.~\eqref{eq:ope} that the leading-power correction term $m_0(\tau)$ multiplies the twist-two matrix elements in the same way independent of their spin. This observation is still
valid when $m_0(\tau)$ plays the additional role to account
for the scheme dependence in regularizing the 
leading renormalon divergence in the coefficient function $C_k(\alpha_s)$. This is because all $C_k(\alpha_s)$ has the exactly the same leading renormalon series as a quark ``pole'' mass. Moreover, this leading renormalon series exponentiates such that it matches exactly the mass renormalization of the Wilson line in the quasi-PDF operator. Therefore, if we renormalize the large-$P^z$ spatial correlators $h^B(z,P^z,a)$ with the $Z^R(z,a,\mu,\tau)$ and $m_0(\tau)$ from the previous section, the resulting $h^R(z,P^z,\mu,\tau)$ can be matched to light-cone PDFs with $\tau$ prescription, e.g., PV, for the leading renormalon in the matching coefficient without any explicit leading power corrections. 

To extract the $x$-dependent PDF with linear accuracy in $1/P_z$, one needs either to match the Euclidean coorelation functions to the lightcone with linear-$z$ accuracy in coordinate space then perform a Fourier transformation, or to first obtain the $x$-dependent quasi-PDF then match it to PDF with the linear-$1/P_z$ accuracy. The former approach faces the problems of the breaking down of twist expansion and the existence of the Landau pole when going beyond $z\sim\Lambda^{-1}_{\rm QCD}$. So we take the second approach, which then requires the LRR correction to the momentum-space matching. In this approach, to avoid the Landau pole, we derive the regularized LRR correction with fixed renormalization scale $\mu$, then apply the renormalization group resummation in momentum space, as shown in Ref.~\cite{Su:2022fiu}. Utilizing the fact that the LRR correction is universal to all $C_k$,
the correction to the momentum-space matching kernel $\mathcal{C}(x,y,\mu,P_z)$ is just the Fourier transformation of the coordinate-space corrections: 
\begin{align}
	\Delta& \mathcal{C}^{\rm LRR}(x,y,\mu,P_z,\tau)=\int \frac{yP_zdz}{2\pi} e^{i (x-y) zP_z}  \nonumber\\
 &\left[\mu z C_k(\alpha_s(\mu),z^2\mu^2)_{\rm PV}-\sum_i\mu z\alpha^{i+1}_s r_i(\mu)\right].
\end{align}
Note that the integrand is linear in $z$, thus is integrated to a singular function which includes derivative of delta functions $\delta'(x-y)$. To avoid the instability of its numerical implementation, it is helpful to introduce a small exponential decay to regularize the long-tail, 
\begin{align}
	z\to z\exp(-\epsilon_m |z|),
\end{align}
where the smaller parameter $\epsilon_m\to0$ along with a finer discretization of the $x\in[0,1]$ region eventually recovers its continuum version. With finite but small $\epsilon_m$, the extra correction introduced by this regularization is of $\mathcal{O}(z^2\epsilon_m\Lambda_{\rm QCD})$, thus is converted to $\mathcal{O}(\epsilon_m\Lambda_{\rm QCD}/(2xP_z)^2)$, not breaking the linear accuracy. In the hybrid scheme~\cite{Ji:2020brr}, where the integral is from $z_s$ to $\infty$, we obtain
\begin{widetext}
\begin{align}
	\Delta& \mathcal{C}^{\rm LRR}(\xi,\mu,p_z,\tau)=N_m\mu\left\{\frac{e^{-\epsilon_m z_s}p_z(1+\epsilon_m z_s+\epsilon_m^2z_s^2)}{\epsilon_m^2\pi}+\frac{1}{\pi}\left(\frac{e^{-\epsilon_m z_s}z_s(\sin[\bar{\xi} z_sp_z])}{\bar{\xi}}\right.\right.\\
	+&\left.\left.\frac{e^{-\epsilon_m z_s}p_z}{(\epsilon_m^2+p_z^2\bar{\xi}^2)^2}\left((\epsilon_m^2-\bar{\xi}^2p_z^2+\epsilon_m^3z_s+\epsilon_mp_z^2\bar{\xi}^2z_s)\cos[\bar{\xi} z_sp_z]-\bar{\xi} p_z(2\epsilon_m+\bar{\xi}^2p_z^2z_s+\epsilon_m^2z_s)\sin[\bar{\xi} z_sp_z]\right)\right)\right\}_+.\nonumber
\end{align}
\end{widetext}
where we have used
\begin{align}
   \xi=x/y,\quad \bar{\xi}\equiv1-\xi,\quad p_z=yP_z,
\end{align}
for simplicity, and a plus function
\begin{align}
    [f(x)]_+=f(x)-\delta(1-x)\int_0^1f(\nu)d\nu
\end{align}
to compensate a neglected $\delta(1-\xi)$ term in this correction.
Testing with some different $\epsilon_m\in[20,100]$ MeV, we find the results are consistent and stable.

\begin{figure}[tbhp!]
    \centering
    \includegraphics[width=0.9\linewidth]
    {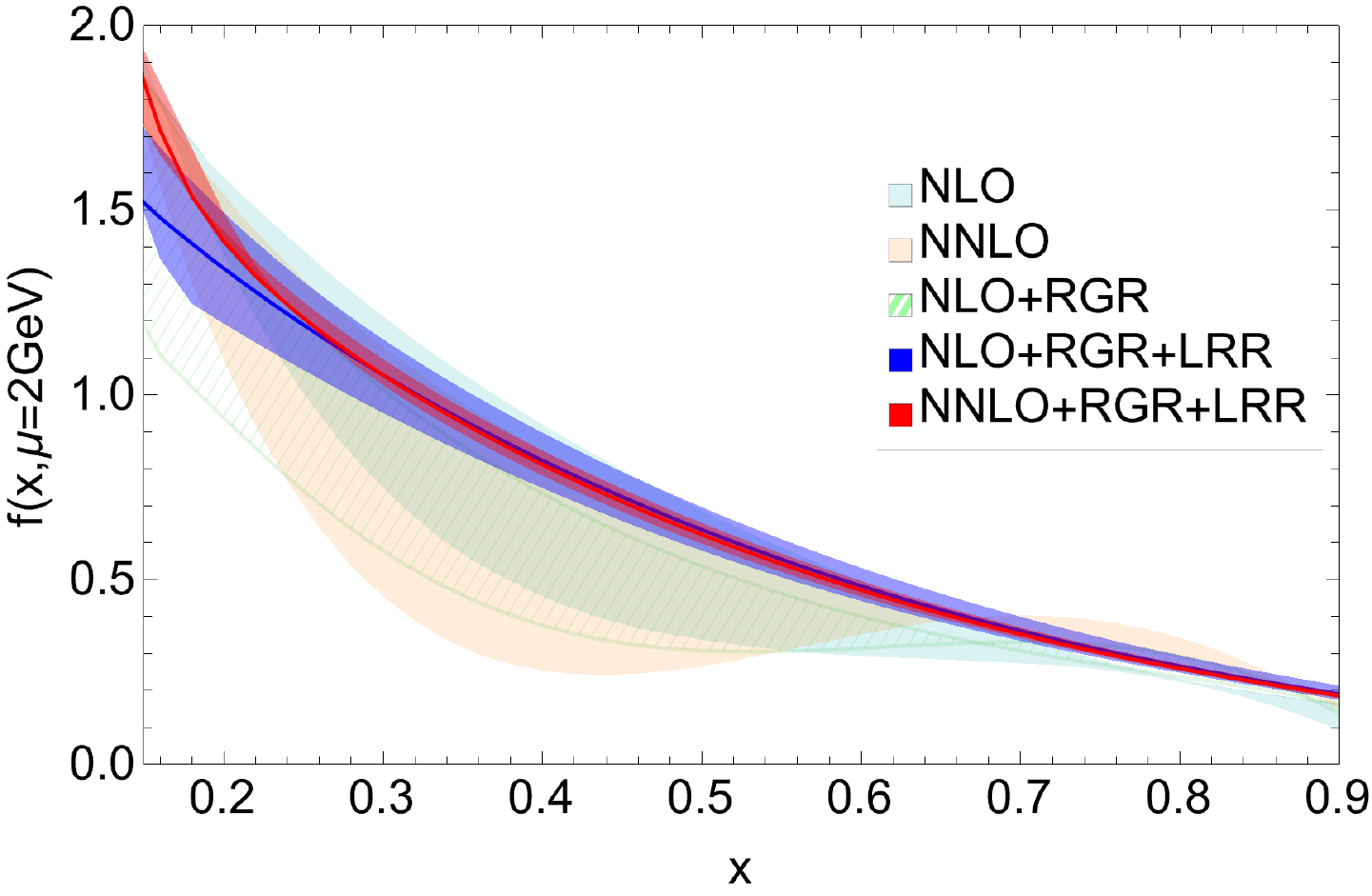}
    \caption{The effect of leading-renormalon resummation (the red and blue band) on the pion PDF, compared with fixed-order results in the background.}\label{fig:pion_example}
\end{figure}

We then apply the leading-renormalon resummed matching coefficients and the corresponding $m_0$ to the analysis of the pion PDF lattice data~\cite{Gao:2021dbh}, with results shown in Fig.~\ref{fig:pion_example}. The results from fixed-order perturbation theory from Fig.~\ref{fig:nlo_compare} are shown again for comparison. The $m_0(\tau)$ used in calculating the blue (red) band is from the bottom plot in Fig.~\ref{fig:c0_compare}. The error bands are obtained by varying the starting point of the RG evolution in both $m_0(\tau)$ extraction and perturbative matching. The results show much reduced error bands from LRR because of the much smaller uncertainty in $m_0(\tau)$. Interestingly, the NNLO+RGR+LRR result suggests a even smaller error after matching, because the scale variation in the RGR matching cancels most of the corresponding $m_0$ difference in coordinate space. Moreover, the consistency in $x>0.2$ between NLO and NNLO suggests good convergence of the perturbation theory after LRR, the same as our observation in coordinate space.

In conclusion, we made a first systematic study of the leading power accuracy in LaMET calculations of PDFs and suggested an approach to make practical progress. We show the importance to define the renormalization scheme for the linear divergence consistently in the renormalization and the perturbative matching. To achieve that, we resum the leading IRR in the the perturbative Wilson coefficients. The leading IRR is universal in all orders of Wilson coefficients. Thus after renormalization in the same scheme, the $P^z=0$ lattice correlators are also consistent with perturbative results up to linear accuracy. Then $P^z>0$ PDF extraction allows the same level of precision, provided a perturbative matching in the same scheme is applied. An application to the pion PDF shows that our approach significantly reduces the uncertainty from linear corrections. This is a necessary step towards the future high-precision calculation of PDFs on lattice.
\section*{Acknowledgement}
We thank Lattice Parton Collaboration (LPC) and ANL-BNL lattice collaboration for sharing their data for this work. We also thank Yong Zhao and Jian-Hui Zhang for useful discussions. This research is supported by the U.S. Department of Energy, Office of Science, Office of Nuclear Physics, under contract number DE-SC0020682. J.H. is partially supported by the Center for Frontier Nuclear Science at Stony Brook University. Y.S. is partially supported by the U.S.~Department of Energy, Office of Science, Office of Nuclear Physics, contract no.~DE-AC02-06CH11357.


\end{document}